\documentclass[AMA,LATO1COL]{WileyNJD-v2} 

\articletype{Research Article}%

\usepackage{amssymb,amsfonts}

\begin{document}
\title{Bayesian uncertainty quantification framework for wake model calibration and validation with historical wind farm power data}

\author[1]{Frederik Aerts}

\author[1]{Luca Lanzilao}

\author[1]{Johan Meyers}

\authormark{Aerts, Lanzilao, and Meyers}

\address{\orgdiv{Department of Mechanical Engineering}, \orgname{KU Leuven}, \orgaddress{\state{Leuven}, \country{Belgium}}}

\corres{Johan Meyers, Department of Mechanical Engineering, KU Leuven, Leuven, Belgium. \\ \email{johan.meyers@kuleuven.be}}

\abstract[Summary]{The expected growth in wind energy capacity requires efficient and accurate models for wind farm layout optimization, control, and annual energy predictions. Although analytical wake models are widely used for these applications, several model components must be better understood to improve their accuracy. To this end, we propose a Bayesian uncertainty quantification framework for physics-guided data-driven model enhancement. The framework incorporates turbulence-related aleatoric uncertainty in historical wind farm data, epistemic uncertainty in the empirical parameters, and systematic uncertainty due to unmodelled physics. We apply the framework to the wake expansion parameterization in the Gaussian wake model and employ historical power data of the Westermost Rough offshore wind farm. We find that the framework successfully distinguishes the three sources of uncertainty in the joint posterior distribution of the parameters.  On the one hand, the framework allows for wake model calibration by selecting the maximum a posteriori estimators for the empirical parameters. On the other hand, it facilitates model validation by separating the measurement error and the model error distribution. In addition, the model adequacy and the effect of unmodelled physics are assessable via the posterior parameter uncertainty and correlations. Consequently, we believe that the Bayesian uncertainty quantification framework can be used to calibrate and validate existing and upcoming physics-guided models.}

\keywords{Bayesian inference, uncertainty quantification, analytical wake model, calibration, validation, historical power data}

\maketitle

\section{Introduction}
Wind energy is expected to play a major role in the energy transition by quadrupling the added annual capacity by 2030 and increasing its global penetration from 5\% to 35--50\%.\cite{IEA} However, the technology still faces some fundamental challenges including the improved understanding of atmospheric and wind power plant flow physics.\cite{Veers} One aspect is the accurate modelling of wind turbine wake effects for wind farm layout optimization, control and reliable annual energy predictions. 

Since the 1980s, several analytical wind turbine wake models have been introduced.\cite{reviewPA,wakemodelsDTU} One of the first of this kind is the Jensen model,\cite{Jensen} which is based on mass conservation and assumes a top-hat shape of the wake. Due to its simplicity, the Jensen model has been extensively used in literature and commercial software.\cite{reviewPA} However, the increase in computational power and the availability of better experimental data have led to the development of alternative engineering models, such as the Larsen \cite{Larsen} and Frandsen model,\cite{Frandsen} and eventually the Gaussian wake model.\cite{Gaussian} In contrast to the Jensen model, the Gaussian model's derivation includes both mass and momentum conservation and assumes a Gaussian profile based on a self-similarity argument. To further improve the Gaussian model in the near wake, the double-Gaussian \cite{doublegaussian} and super-Gaussian \cite{supergaussian} were developed. In recent years,  purely data-driven models have also emerged, favoured by the large amounts of available simulation and historical wind farm data.\cite{reviewDATA} However, their interpretability and generalizability have been questioned and the usage of data for the development of physics-guided models is often recommended.\cite{reviewDATA}

To predict the generated power of a wind farm, the interaction of wind turbine wakes has to be taken into account. For this purpose, wake-merging methods were developed which allow to compute the complete flow field inside the wind farm. The inlet velocity of every turbine is then derived from the flow field.  Afterwards, it is translated into power by using the turbine power curve. The most common wake-merging methods are based on a superposition of velocity \cite{Lissaman}\cite{Niayifar} or energy deficits.\cite{Katic,Voutsinas} More recently, Lanzilao and Meyers introduced a new wake-merging method,\cite{Lanzilao} in which a recursive method is used to take into account heterogeneous background velocity fields caused by, for instance, coastal gradients.\cite{coastalgradients}

Wind turbine wakes expand due to the flow mixing at the edges of the wake. This wake expansion is in most of the wake models represented by a linear relationship between the wake width and the distance from the rotor with a rate $k^\ast$ denoting the wake expansion parameter. However, the wake expansion is still considered a sensitive and not fully understood model component.\cite{anamerge} The Jensen model considers this parameter to be fixed, with a value of $0.04$ or $0.05$ for off-shore \cite{k_Barthelmie}  and $0.075$  for on-shore  \cite{k_Gocmen} wind turbines. As a higher turbulence intensity results in faster wake recovery and lower power deficits due to enhanced flow mixing,\cite{Hansen} Niayifar\cite{Niayifar} fitted a linear relationship between the wake expansion and the local turbulence intensity within the Gaussian wake model. 

Although multiple studies have  validated analytical wake models with historical power data, they do not fully account for the uncertainty present in the data yet.\cite{Doekemeijer} However, the large uncertainty on the input of the wake models can cause a lack of evidence of model adequacy, with as a result that no clear improvement of the predictive capabilities of wake models of higher fidelity can be found. \cite{MurciaPhD} On the other hand, the parameter uncertainty is usually not assessable in model calibration\cite{Markfort,learning_SCADA} and empirical model development.\cite{Ishihara}  Especially in historical data, the uncertainty due to turbulence and uncaptured atmospheric physics is considerable. For instance, the linear relationship between the wake expansion rate and the turbulence intensity was originally fitted against three turbulence intensities values from LESs. Later, Fuertes\cite{Fuertes} used measurements by nacelle-mounted wind lidars to estimate this relationship as well as another empirical parameter, namely the initial wake width. Eventually, Markfort\cite{Markfort} calibrated the linear model with Supervisory Control And Data Acquisition (SCADA) data of an onshore wind farm. Although the three studies show the same positive correlation between the wake expansion and the turbulence intensity, the coefficients that were found, are quite different. Nowadays, it is common practice to use the original fitting rule proposed by Niayifar and Porté-Agel (e.g. in FLORIS\cite{FLORIS}). 

In this article, we propose a Bayesian uncertainty quantification framework for (1) model validation by distinguishing measurement errors arising from turbulent fluctuations and the model error (including its bias), and (2) physics-guided data-driven model enhancement by calibrating empirical parameters and interpreting their posterior probability density functions. The Bayesian inference is widely used for parameter uncertainty quantification in physical models \cite{Toussaint,RBC,Zhang,MurciaPhD} and is considered to be a promising approach to data-driven, yet physics-guided and interpretable model enhancement.\cite{natureBI} The framework will be applied to the wake growth parameterization because of its uncertainty and sensitivity. The adopted analytical wake model consists of  the the Gaussian wake model because of its wide acceptance \cite{Doekemeijer,FLORIS} combined with Lanzilao's wake-merging method for its generalizability to heterogeneous background fields.\cite{Lanzilao} The windfarm under analysis is the Westermost Rough farm.

The article is structured as follows: Section \ref{sec:methods} outlines the analytical model for the wind farm power output and the data preprocessing procedure. In Section \ref{sec:bi_framework}, the Bayesian framework is constructed by examining the aleatoric uncertainty due to turbulent fluctuations, the epistemic uncertainty in the empirical model parameters, and the systematic uncertainty due to unmodelled physics. Section \ref{sec:res&dis} presents the joint posterior probability distribution of all parameters and applies the framework to calibration and validation of two wake expansion rate parameterizations. Finally, a summary and a discussion of future research possibilities are given in Section \ref{sec:conclusion}.

\begin{figure} \centering
\includegraphics[width=0.5\linewidth]{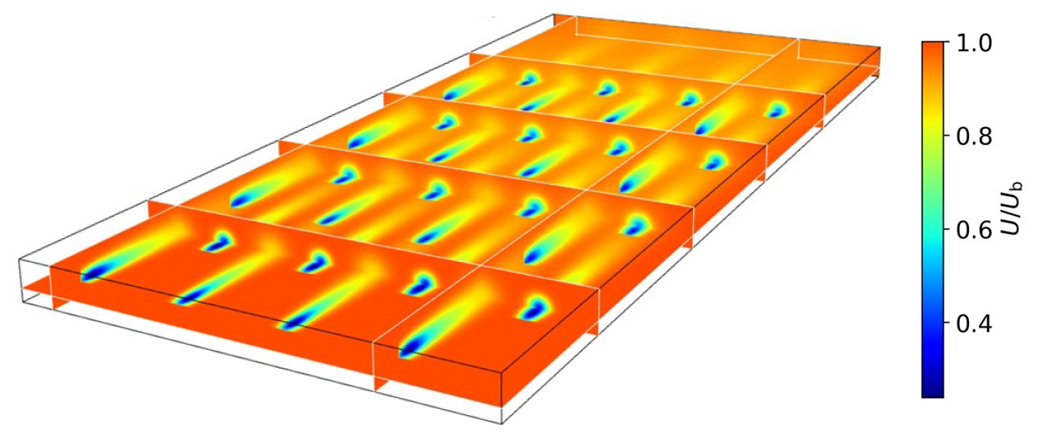}
\caption{Velocity field obtained with an analytical wake model consisting of Gaussian wake model and recursive wake-merging method for a staggered wind farm with 32 turbines with rotor diameter $D=154$ m, streamwise and spanwise turbine spacings of $6D$, thrust coefficient $C_T = 0.85$, upstream turbulence intensity $I_\infty = 12\%$ and background velocity $U_b = 10$ m/s. Figure adapted from Lanzilao and Meyers,\cite{Lanzilao} Wind Energy 25, 237-259.}\label{fig:velfarm}
\end{figure}

\section{Methods}\label{sec:methods}
In Section \ref{sec:anamodel}, the analytical wake model adopted in this article is presented. In addition, the data and preprocessing procedure are discussed in Section \ref{sec:data}.

\subsection{Analytical wake model}\label{sec:anamodel}
The normalized velocity deficit in the wake $W_i(\boldsymbol{x})$ of turbine $i$ is modelled as the product of the maximum velocity deficit compared to the background field $U_b(\boldsymbol{x})$ at each downstream location and a shape function. Here $\boldsymbol{x} = (x,y,z)$ represents a coordinate system with $x$ {normal to the rotor} and along the streamwise direction, $y$ {parallel to the rotor}, and $z$ in the vertical direction. In this frame of reference, $\boldsymbol{x}_i = (x_i,y_i,z_i)$ denotes the $i$-th turbine hub coordinates. The Gaussian wake model then states that
\begin{equation}
   W_i(\boldsymbol{x}) = \frac{U(\boldsymbol{x}) - U_b(\boldsymbol{x})}{U_b(\boldsymbol{x})}
 = \left( 1 - \sqrt{1-\frac{C_{T,i}}{8(\sigma/D_i)^2}} \right)  \exp\left( -\frac{(y-y_i)^2+(z-z_{i})^2}{2  \sigma^2} \right) ,
\end{equation}
for $x > x_i$.  Further, $C_{T,i}$ is the wind turbine thrust coefficient and $D_i$ is the turbine rotor diameter which will equal $D$ for every turbine in the current study. The wake width, which is represented by $\sigma$, is assumed to grow linearly with the distance from the rotor $x-x_i$ as
\begin{equation}
\frac{\sigma}{D_i} = k^* \frac{x-x_i}{D_i} + \varepsilon,
\end{equation}
where $k^*$ is the wake expansion coefficient. The initial wake width $\varepsilon D$ is empirically related to\cite{Gaussian} $\beta$ which is related to the induction in the actuator disk model as\cite{WEhandbook}  
\begin{equation}
    \varepsilon = 0.2 \sqrt{\beta} \text{\quad with \quad} \beta = \frac{1}{2} \frac{1+\sqrt{1-C_{T,i}}}{\sqrt{1-C_{T,i}}}.
\end{equation}

The original linear relationship that Niayifar fitted to include the faster wake recovery with increased turbulence intensity is given as
\begin{equation}\label{eq:kTI}
    k^* = a I+  b = 0.3837 I + 0.003678.
\end{equation}
As the local turbulence intensity is mainly dependent on the wind turbine directly upwind,\cite{Frandsen}  Niayifar proposed the local turbulence intensity at turbine $j$ to equal 
\begin{equation}
I_j^2 = I_\infty^2 + \max_i\left(\frac{A_w}{A} I_i^+\right)^2,
\end{equation} 
with $I_\infty$ the upstream turbulence intensity and $A_w$ the overlapping area of the wake of turbine $i$ with diameter $4\sigma$ and the rotor area $A$. Here, the added turbulence intensity $I_i^+$ is computed with the Crespo and Hernandez model.\cite{Crespo}

In the case of uni-directional flow (which is used in the current manuscript), the wake-merging method proposed by Lanzilao and Meyers uses a self-similarity argument to arrive at the following formula 
\begin{equation}
U(\boldsymbol{x}) = U_b(\boldsymbol{x}) \prod_{i=1}^{N_T} \left[1- W_i(\boldsymbol{x}) \right],
\label{eq:wakemerging}
\end{equation}
with $N_T$ the total number of turbines (note that a more general formula exists for cases that are not unidirectional, but this is not needed in the current work). The inlet velocity $U_i$ at turbine $i$ is then computed via a quadrature rule over the rotor area. Using the manufacturer's thrust curve and power curve, the thrust and power of turbine $i$ is then found. An example of $U(\boldsymbol{x})$ using 32 turbines with $D=154$ m is reported in Fig. \ref{fig:velfarm}. 

In what follows, we refer to the analytical wake model as
\begin{equation}
\widetilde{\boldsymbol{P}} = \mathcal{M}(\boldsymbol{\vartheta}_e, \boldsymbol{\varphi}), \label{eq:wakemodel}
\end{equation}
where $\widetilde{\boldsymbol{P}} = [\widetilde{P}_1, \ldots, \widetilde{P}_{N_T}]^T$ is a vector with the modelled normalized power $\widetilde{P}_i$ of each turbine  with respect to the power output of an undisturbed upstream turbine. As only uniform background velocity fields will be considered, this corresponds to the power obtained from the power curve for the free stream speed. Similarly, $\mathcal{M}$ represents the analytical model with $\boldsymbol{\vartheta}_e$ the set of empirical parameters of the wake expansion model (e.g. $k^*$ or $a,b$ in Eq. \ref{eq:kTI}) and $\boldsymbol{\varphi}$ the physical parameters, such as the background velocity field $U_b(\boldsymbol{x})$, the wind direction, the upstream turbulence intensity $I_\infty$, and the layout of the wind farm. Note that, because of model errors, $\widetilde{\boldsymbol{P}}$ will be different from the true power $\boldsymbol{\mathcal{P}}$ (see Section \ref{sec:sources_uncertainty} for more details).

\subsection{Data and preprocessing procedure}\label{sec:data}
\begin{figure}[t] \centering
\includegraphics[width=0.5\linewidth]{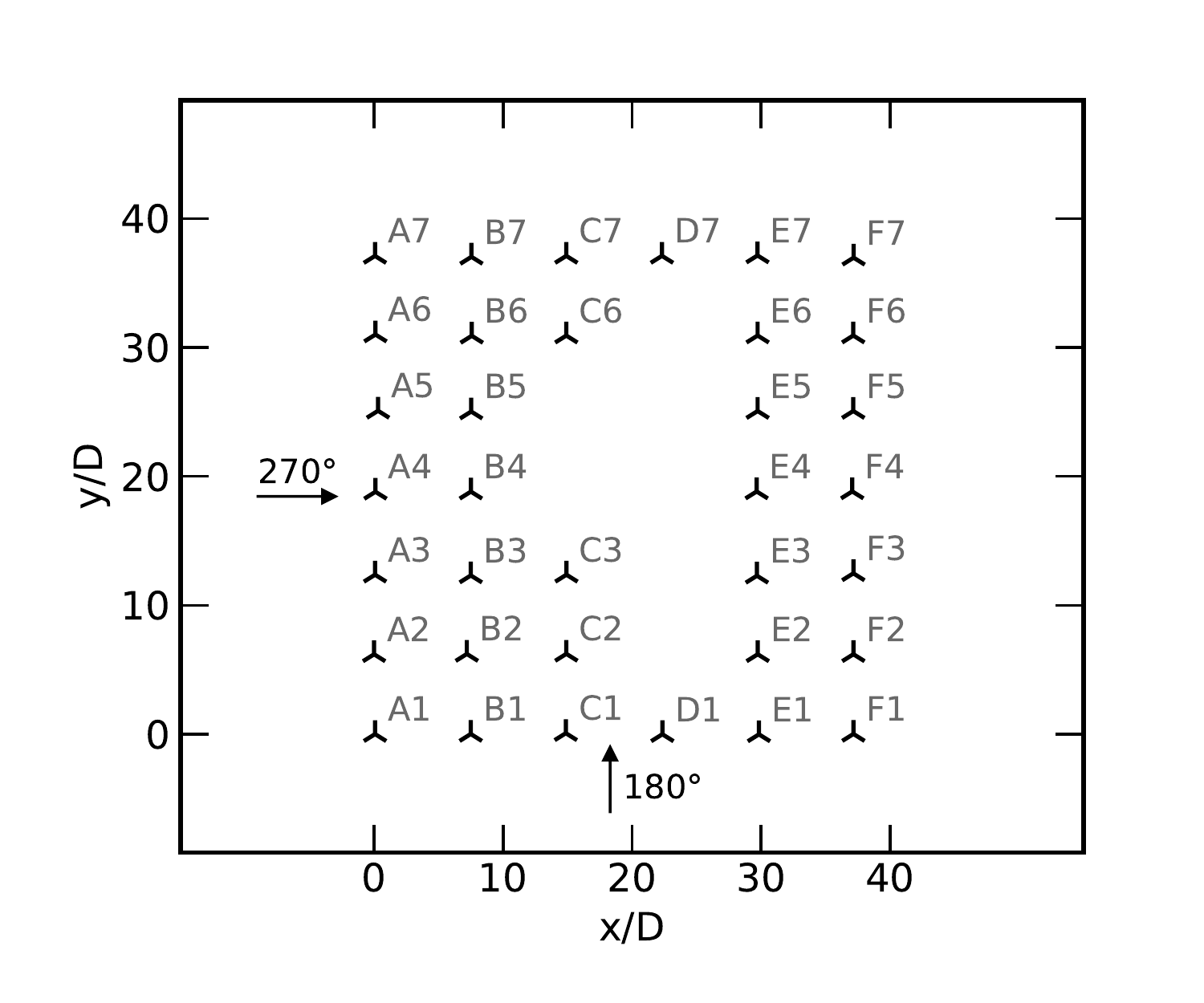}
\caption{Layout of the Westermost Rough wind farm with the turbine names and an indication of the wind directions that will be used for uncertainty quantification.}\label{fig:WMR_layout}
\end{figure}

The power data set used in this article comes from historical 10 minute averaged SCADA data of the Westermost Rough wind farm from July 2016 to December 2017. 
The farm consists of 35 SWT 6MW turbines with a rotor diameter of $D=$ 154 m in a configuration as shown in Fig. \ref{fig:WMR_layout}. Because of the open space in the middle of the farm, its geometry is of interest for the study of wake recovery. 
In the current work, we focus on cases with a wind direction of $270\pm 2.5^\circ$ and $180\pm 2.5^\circ$ to demonstrate the framework.

As the SCADA data contain faulty sensor measurements and periods of downtime due to maintenance, a cleaning procedure based on the following criteria has been performed:\cite{Doekemeijer}
\begin{enumerate}
\item \textit{normal operation}: All power measurements must be strictly positive and cannot exceed the rated power. 
\item \textit{all turbines active}: The turbine must operate the whole time period, i.e. 600 s.
\item \textit{consistent measurement error}: The measured wind speed and power must be consistent with the manufacturer's power curve within reasonable bounds (see below).
\end{enumerate}
Figure \ref{fig:data_doekemeijer} summarises the data cleaning procedure. Based on criterion three, instances with a measured turbine inlet speed and power output that fall outside Doekemeijer's \cite{Doekemeijer} bounds around the manufacturer's power curve are rejected. 
In addition, the wind speed is limited to 12 m/s to make the power curve invertible for inference. Moreover, since the power is constant for higher wind speeds, there is no reason to apply Bayesian inference in this region. Lastly, time stamps with a normalized farm power greater than one are excluded as these may point to the presence of a very strong coastal gradient or non-stationary situations such as a front being convected through the farm.

\begin{figure}[t] \centering
\includegraphics[width=0.5\linewidth]{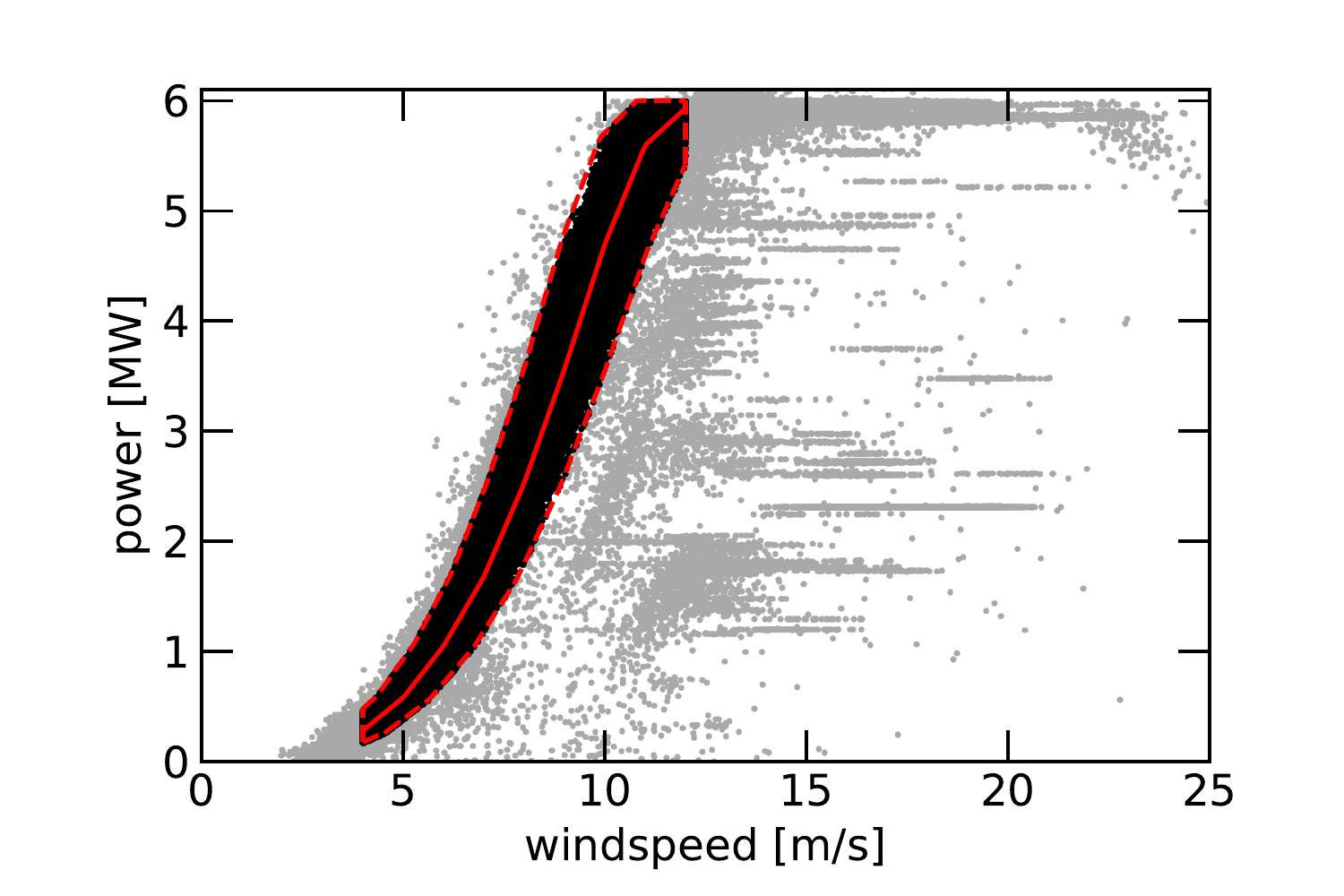}
\caption{Data cleaning procedure on turbine level based on the power curve with the red solid curve representing the manufacturer's power curve and the dashed red curves denoting Doekemeijer's acceptance bounds.\cite{Doekemeijer} Black points are accepted whereas grey points are rejected.} \label{fig:data_doekemeijer}
\end{figure}

The wind direction for each timestamp is estimated as the median yaw angle of all turbines in normal operation. Although the static yaw misalignment is negligible, the sensor signals are often offset from the true orientation of the turbines. For this reason, the median is used to limit the influence of the most extreme outliers. In addition, the residual offset must be corrected via northing calibration.\cite{Nygaard2022,Doekemeijer} To this end, we follow Nygaard \& Hansen,\cite{Nygaard2016} and use a constant yaw offset of $-1.88^\circ$ for the data period that we consider (July 2016 -- December 2017). In what follows, this main wind direction is also denoted as the streamwise direction. The background speed is taken as the average of the inlet speeds of the upstream turbines via the power curve, and is presumed constant throughout the farm ($U_b(\boldsymbol{x})=U_\infty$ in Eq.~\ref{eq:wakemerging}). In this manner, blockage and wakes of neighbouring wind farms are not considered.
The streamwise turbulence intensity is computed as the average of the upstream turbines' turbulence intensity which is calculated as the standard deviation during the 10 min interval of the anemometer wind speed measurements at hub height divided by the average speed during the interval.

\section{Bayesian uncertainty quantification framework}\label{sec:bi_framework} 
In this section the Bayesian uncertainty quantification framework is constructed. In Section \ref{sec:sources_uncertainty}, we identify the sources of uncertainty in the context of analytical wake models and SCADA data in a similar manner as Kennedy and O'Hagan.\cite{O'Hagan} Section \ref{sec:biformulation} brings all sources of uncertainties together in a Bayesian uncertainty quantification framework. Section  \ref{sec:applicationsBI} discusses the application of this framework to calibration and validation. Finally, Section \ref{sec:validation_biframework} shows how to validate the Bayesian framework itself. 

\subsection{Sources of uncertainty}\label{sec:sources_uncertainty}
Wake models focus on the prediction of the mean power of each turbine in a farm, given steady atmospheric background conditions. Let us consider the true power 
\begin{equation}
\boldsymbol{\mathcal{P}} = \mathcal{T}(\boldsymbol{\psi}),
\end{equation}
where $\mathcal{T}$ is the true process, and $\boldsymbol{\psi}$ a general vector of all relevant physical conditions that describe the process (e.g., the exact background velocity field, atmospheric stability, yaw turbine set points,  turbulent flow structures,  wind veer, ...). Furthermore $\boldsymbol{\mathcal{P}} = [P_1, \ldots, P_{N_T}]^T$ is a vector containing the power output per turbine, given conditions $\boldsymbol{\psi}$, and normalized with the power of an `undisturbed' upstream turbine (cf. \S\ref{sec:data} above for its practical definition).

Unfortunately, the true power is unknown, as any measurement of the effective power output will introduce errors. These errors can be caused by the precision of the sensors used, but in the current work, we assume that sensor errors are dominated by errors that arise from imperfect averaging. In reality, meteorological conditions are slowly changing, and to be able to presume `constant' background conditions, we can not perform averaging of the measured turbine power over too long time intervals. In practise, we will use the conventional 10-minute averaged power, leading to a measured power  $\boldsymbol{P}^*$ that is different from the true (ensemble averaged) steady power $\boldsymbol{\mathcal{P}}$. This is further discussed in \S\ref{sec:measurementerror}, where we propose a parameterized formulation of this error.

Also the true process $\mathcal{T}$ is unknown. The best substitute we have is the model $\mathcal{M}(\boldsymbol{\vartheta}_e, \boldsymbol{\varphi})$. This introduces two additional sources of uncertainty. First of all, the empirical parameters $\boldsymbol{\vartheta}_e$ are a priori unknown, or estimated from a limited amount of data, and are thus subject to epistemic uncertainty. Moreover, the model $\mathcal{M}$ may introduce additional model errors, resulting from a simplified representation of the physics, and from the fact that the model may only be using a subset $\boldsymbol{\varphi}$ of all relevant physical conditions $\boldsymbol{\psi}$ that matter for the true process. This is further discussed in \S\ref{sec:unmodelledphysics}.

\subsubsection{Turbulent fluctuations}\label{sec:measurementerror}
Due to the turbulent nature of the background velocity field, the 10-minute averaged measurements of the wind turbine power outputs will not equal the assumed steady state value $\boldsymbol{\mathcal{P}}$. Based on central limit theorem, the 10-minute averaged measurements are considered normally distributed with $\boldsymbol{\mathcal{P}}$ as mean. Faulty sensor measurements are considered to be negligible compared to the turbulent fluctuations after preprocessing the data. Thus, the assumed true steady state turbine powers $\boldsymbol{\mathcal{P}}$ can be related to the measurements $\boldsymbol{P}^*$ as
\begin{equation}
\boldsymbol{P}^* = \mathcal{T}(\boldsymbol{\psi}) + \mathcal{E}_T(\boldsymbol{\psi}),
\end{equation}
where $\mathcal{E}_T(\boldsymbol{\psi})$ is the measurement error introduced by the imperfect averaging of the turbulent fluctuations in the wind. The fluctuations on the power of different wind turbines will be correlated due to the convection of turbulent structures through the farm. Proper orthogonal decomposition shows that these structures are primarily convected along the streamwise direction, following the background velocity field.\cite{POD} Correlation studies in LES and real farm data have also shown that these correlations are mainly streamwise due to the movement of turbulent structures in the farm.\cite{Stevens,Andersen} Moreover, Lukassen derived an analytical expression for the spatio-temporal correlations in wind farms.\cite{anacorr} Based on this work, we propose a Gaussian spatial decorrelation of the power outputs of the wind turbines, i.e.
\begin{align}
 \mathcal{E}_T|\boldsymbol{\psi}  &\sim \mathcal{N}(0, \boldsymbol{\Sigma}_T), \label{eq:measerror}\\
 {\Sigma}_{T,ij}  =\sigma_{i}\sigma_{j} \exp &\left[- \left(\frac{s_i - s_j}{L_s} \right)^2 - \left( \frac{n_i - n_j}{L_n} \right) ^2 \right] , \label{eq:measerrorSigma}
\end{align}
where $(s_i,n_i)$ are the streamwise and spanwise coordinates of the $i$-th turbine. In addition, $L_s$ and $L_n$ are the streamwise and spanwise correlation length scales, and $\sigma_i$ is the standard deviation on the measured 10-minute averaged power output of turbine $i$. 

In the current work, we will make the simplifying assumption that $\mathcal{E}_T|\boldsymbol{\psi} \sim \mathcal{E}_T|\boldsymbol{\varphi}$, i.e. our error model is not dependent on possible unrecognized physics. Moreover, we further assume that the standard deviation $\sigma_i$ is identical and equal to $\sigma_T$ for each turbine $i$ after normalizing w.r.t. the power output of the upstream turbines. Thereby the effect of the increased turbulence intensity at lower wind speeds is neglected. In addition, we set $L_{n}$ to $D$ based on the assumption that there is no significant correlation over spanwise turbine distances. This finally leaves two parameters  $\boldsymbol{\vartheta}_t = \{\sigma_T, L_s\}$ that will be estimated in the Bayesian framework (see \S\ref{sec:biformulation}). It should be straightforward to extend this parameter set (e.g. including an estimation of $L_{n}$, dependence on atmospheric stability, etc.). This will typically require inference over larger data sets, and is an interesting topic of future research, but is not further considered in the current manuscript.

Finally, since the fluctuations on the power are caused by the turbulent fluctuations in the wind farm flow field, it should be mentioned that the conditions $\boldsymbol{\varphi}$ also contain uncertainty. The uncertainty on the background wind speed $U_b$ is incorporated in the power fluctuations of the upstream turbines as $U_b$ is estimated via the power curve. The uncertainty in the wind direction is incorporated by binning the data in bins of 5 degrees. If the bin width is too large, different physical scenarios of alignment with different wake losses cannot be distinguished in model validation.\cite{Doekemeijer}

\subsubsection{Unmodelled physics}\label{sec:unmodelledphysics}
The engineering model $\mathcal{M}(\boldsymbol{\vartheta}_e, \boldsymbol{\varphi})$ only considers a subset $\boldsymbol{\varphi}$ of all relevant conditions $\boldsymbol{\psi}$ (e.g. stability not considered, sea surface roughness, a lack of resolution/detail in representing velocity fields, ...). 
Consequently, a model error $\mathcal{E}_B$ must be present, which will vary for different unrecognised conditions $\boldsymbol{\psi'}=\boldsymbol{\psi} \setminus \boldsymbol{\varphi}$ in the engineering model. Not including the model bias explicitly in the Bayesian framework inevitably results in an incorrect inference of the turbulent fluctuations, as these will then represent both the fluctuations in the data and the model bias.\cite{Gelman} Because the turbulent fluctuations have to be the same for every model, whereas the model bias serves as a metric for model validation, the distinction between the two is important. 

Ghanem et al.\cite{Ghanem} suggest including the model uncertainty in the distribution of the empirical parameter and propose a methodology in the absence of measurement errors. However, they admit that low-dimensional models will not be able to reach all possible outputs by varying the empirical parameters. Therefore, this approach may again lead to an inadequate inference of the model and measurement error. As this is the case for the constant wake expansion rate model, this method is not further explored. For this reason, we propose an additive model error per wind turbine $\mathcal{E}_B$. As the model is low-dimensional and does not capture all physics given $\boldsymbol{\varphi}$, the error will again depend on all conditions $\boldsymbol{\psi}$ and is given by
\begin{equation}
\boldsymbol{\mathcal{P}} = \mathcal{T}(\boldsymbol{\psi}) \triangleq \mathcal{M}(\boldsymbol{\vartheta}_e, \boldsymbol{\varphi}) + \boldsymbol{E}_B(\boldsymbol{\psi}),
\end{equation}
with $\boldsymbol{E}_B=[\varepsilon_1, \ldots , \varepsilon_{N_T}]^T$ the normalized power errors at each turbine. Formally, the model error is deterministic, i.e. $\mathbb{P}(\mathcal{E}_B|\boldsymbol{\psi}) = \delta(\mathcal{E}_B-\boldsymbol{E}_B(\boldsymbol{\psi}))$, 
with $\delta(\cdot)$ the Dirac-delta function. In practise, however, we do not know the unmodelled physics $\boldsymbol{\psi'}$, so the best we can achieve is estimating the distribution 
\begin{equation}
\mathbb{P}(\mathcal{E}_B|\boldsymbol{\varphi})=\int  \delta(\mathcal{E}_B-\boldsymbol{E}_B(\boldsymbol{\psi})) \mathbb{P}(\boldsymbol{\psi'}) {\rm d}\boldsymbol{\psi'}. \label{eq:modelerrordef}
\end{equation}
Here, we will make the strong assumption that the model error is multivariate normally distributed with a bias term $\boldsymbol\mu_B=[\mu_{1}, \cdots, \mu_{N_T}]^T$ (i.e. one per turbine), and $\boldsymbol{\Sigma}_B \in \mathbb{R}^{N_T \times N_T}$ the covariance matrix. This could be justified when the model bias is the result of many additive errors due to different unrecognised conditions. In addition, if we are only interested in the means, variances and covariances, then a multivariate normal distribution is the maximum entropy distribution.\cite{McElreath} Thus, we represent the model error as
\begin{equation}
\mathcal{E}_B|\boldsymbol{\varphi} \sim \mathcal{N}(\boldsymbol\mu_B, \boldsymbol{\Sigma}_B). \label{eq:modelerrordist}
\end{equation}

In the following, we further simplify the model error parametrization to arrive at a form that can be estimated in practise and yields a parametrization that is independent of the precise farm layout or the number of turbines.
Defining a bias term for every turbine is not easily generalizable to multiple wind directions (as the bias would be direction-dependent), nor to farms with different amounts of turbines. Moreover, inferring all individual bias terms would require a lot of data, and could still result in higher uncertainty on the empirical parameters as more combinations of parameters lead to the same result. Therefore, we propose to use a lower-dimensional representation of the model error. We presume that in wake models a similar bias is expected for turbines with the same amount of affecting upstream turbines. To this end, let us define for each turbine $i$, the number of wakes $\zeta_i$ it is affected by, with a disk averaged wake deficit larger than $1\%$. For instance, for the Westermost Rough layout and the wind direction of 270$^\circ$ in Fig. \ref{fig:WMR_layout}, $\zeta_{E1} = 4$, $\zeta_{E4} = 2$, etc. We then propose to express the mean and covariance in Eq.~\ref{eq:modelerrordist} as
\begin{align}
\mu_{i} &= \delta_{\zeta_i}\\
 {\Sigma}_{B,ij}  &= 
 		\begin{cases}
            \sigma_{\zeta_i}\sigma_{\zeta_j}  \quad &  \zeta_i = \zeta_j \\ 
            0  \quad  &   \zeta_i \neq \zeta_j 
        \end{cases} \label{eq:modelSigma}
\end{align}
For the layout in Fig. \ref{fig:WMR_layout} and the $270^\circ$ case, this leads to a parametrization of the model error in 6 bias terms $\boldsymbol{\delta}=[\delta_0, \delta_1, \cdots, \delta_5]$, and 6 respective variances; while the  $180^\circ$ case leads to 7 biases and variances.

Finally, we expect that calibration of wake parameters will at least allow to arrive at a zero farm power bias (in the current work, $\boldsymbol{\vartheta_e}$  has either one or two degrees of freedom $k^*$ or $a,b$, see \S\ref{sec:anamodel}). This can be included in the model error as it is a form of prior knowledge. Here, we impose this by requiring that
\begin{equation}\label{eq:constraint_farmpower}
\sum_i \mu_i = \sum_{\zeta} n_\zeta \delta_\zeta = 0,
\end{equation}
with $n_\zeta$ the number of turbines that are influenced by $\zeta$ wakes. We insert this relation in the Bayesian framework below by using it to simply eliminate one of the parameters in $\boldsymbol{\delta}$.

\subsection{Bayesian formulation}\label{sec:biformulation}
Bayesian inference uses Bayes' theorem to estimate parameters $\boldsymbol{\vartheta}$ from data $\mathcal{D}$ by building the joint posterior probability distribution of $\boldsymbol{\vartheta}$. To this end, a Bayesian model $\mathbb{M}$ must be specified. In the current work, the model $\mathbb{M}$ includes the analytical wake model $\mathcal{M}$ (Eq.~\ref{eq:wakemodel}),  but also the parametrizations of the measurement error (Eq.~\ref{eq:measerror}, \ref{eq:measerrorSigma}) and the model error (Eq.~\ref{eq:modelerrordist}--\ref{eq:modelSigma}). The Bayesian model depends on a parameter set $\boldsymbol{\vartheta}  = \{\boldsymbol{\vartheta}_e, \boldsymbol{\vartheta}_t, \boldsymbol{\vartheta}_b \}$. Here (see also above), $\boldsymbol{\vartheta}_e$ represents the empirical model parameters in the analytical wake model. Hence, $\boldsymbol{\vartheta}_e = \{k^* \}$ or $\boldsymbol{\vartheta}_e = \{a,b \}$ depending on the wake expansion rate parameterisation adopted. The parameters for the turbulent fluctuations are $\boldsymbol{\vartheta}_t = \{\sigma_T, L_s\}$, and the model error parameters are  $\boldsymbol{\vartheta}_b = \{\delta_{\zeta_i}, \sigma_{\zeta_i} \}$. The parameters in the subset $\{ \boldsymbol{\vartheta}_t, \boldsymbol{\vartheta}_b \}$ are also called statistical parameters.

Bayes' theorem can now be used to express
\begin{equation}
\mathbb{P}(\boldsymbol{\vartheta}  | \mathcal{D}, \mathbb{M}) = \frac{\mathbb{P}(\mathcal{D} | \boldsymbol{\vartheta} , \mathbb{M}) \mathbb{P}(\boldsymbol{\vartheta}  | \mathbb{M})}{\mathbb{P}(\mathcal{D}|\mathbb{M})},
\end{equation}
with $\mathbb{P}(\boldsymbol{\vartheta}  | \mathbb{M})$ the \textsl{prior}, $\mathbb{P}(\mathcal{D} | \boldsymbol{\vartheta} , \mathbb{M})$ the \textsl{likelihood}, $\mathbb{P}(\mathcal{D}|\mathbb{M})$ the \textsl{evidence}, and $\mathbb{P}(\boldsymbol{\vartheta}  | \mathcal{D}, \mathbb{M})$ the \textsl{posterior}. Since the evidence is independent of the parameters $\boldsymbol{\vartheta}$, it can be considered a normalizing factor. Consequently, the Bayesian model is fully specified by the likelihood and the prior. The fact that all probabilities are conditioned on the Bayesian model $\mathbb{M}$ will further not be mentioned explicitly, unless necessary.

The prior is the joint probability density of all parameters $\boldsymbol{\vartheta}$. If all empirical and statistical parameters are a priori assumed to be independent, the joint prior distribution takes the following form
\begin{equation}
\mathbb{P}(\boldsymbol{\vartheta} ) = \prod_{i=1}^{N_e} \mathbb{P}( \vartheta_{e,i}) \prod_{j=1}^{N_t} \mathbb{P}( \vartheta_{t,j}) \prod_{k=1}^{N_b} \mathbb{P}( \vartheta_{b,k}).
\end{equation}
In nonlinear models, the empirical model parameters are usually given weakly informative uniform priors.  To this end, a non-informative uniform distribution over the real number line is limited to an interval $[l_i, u_i]$ in order to exclude non-physical values.\cite{Gelman,RBC} Thus,
\begin{equation}
    \vartheta_{e,i} \sim \mathcal{U}(l_i, u_i).
\end{equation}
The statistical parameters require prior distributions as well. The standard deviation of the turbulent power fluctuations must be positive by definition. As the power outputs are normalized, values greater than 0.5 would imply that less than 95\% of the power samples lie in the interval $[0,2]$, which would be rather non-physical. Therefore, a uniform prior $\sigma \sim \mathcal{U}(0,0.5)$ is given. For the length scales, an exponential distribution,\cite{McElreath} $\mathbb{P}(L_s)=\lambda \exp(-\lambda L_s/D)$, with  scale parameter $\lambda = 0.2$ is taken  such that the mean and standard deviation equal 5 turbine diameters and the probability density function dies out exponentially. The normalized bias terms $\delta_\zeta$ are given uniform priors bounded to $[-1,1]$. The standard deviations of the bias terms must be positive and cannot exceed $1$. Therefore, uniform priors $\sigma_{\zeta_i} \sim \mathcal{U}(0,1)$ are given. A summary of the prior distributions is given in Table \ref{tab:priors}. As the influence of the prior decays with more data, Jeffrey's priors are more appropriate for very small datasets.\cite{Toussaint,Gelman}

The data set $\mathcal{D} = \{(\boldsymbol{P}^*_1, \boldsymbol{\varphi}^*_1), \ldots, (\boldsymbol{P}^*_{N_s},\boldsymbol{\varphi}^*_{N_s})\}$ contains $N_s$ samples with $N_s$ the number of 10-minute periods after data cleaning and within the wind direction sector. The vectors $\boldsymbol{P}^*_n $ represent the measured power output of each turbine normalised with respect to the power output of the upstream turbines and $ \boldsymbol{\varphi}^*_n$ are the corresponding measurements of physical parameters.  Given all Bayesian model parameters $\boldsymbol{\vartheta}$, the power measurements $\boldsymbol{P}^*_n$ equal the sum of the deterministic model vector $\mathcal{M}(\boldsymbol{\vartheta}_e, \boldsymbol{\varphi}_n^*)$, the zero-centered turbulent fluctuation term $\mathcal{E}_T$, and the model error $\mathcal{E}_B$. Therefore, the measurements are distributed as
\begin{align}
\boldsymbol{P}_n^* |\boldsymbol{\vartheta},\boldsymbol{\varphi}_n^*  &\sim  \{\mathcal{M}+\mathcal{E}_T+\mathcal{E}_B\} |\boldsymbol{\vartheta},\boldsymbol{\varphi}_n^*  \nonumber \\ 
&\sim \mathcal{N}(\mathcal{M}(\boldsymbol{\vartheta}_e, \boldsymbol{\varphi}_n^*) + \boldsymbol\mu_B, \boldsymbol\Sigma_T + \boldsymbol\Sigma_B). \label{eq:likelihood}
\end{align}
Under the assumption that the samples are independent and identically distributed, the likelihood can be written as
\begin{equation}
\mathbb{P}(\mathcal{D}| \boldsymbol{\vartheta}) = \prod_{n=1}^{N_s} \mathbb{P}(\boldsymbol{P}^*_n | \boldsymbol{\vartheta},  \boldsymbol{\varphi}_n^*) .
\end{equation}
Note that in fact, the samples stem from a time series, so that consecutive samples may not be entirely independent and typically positively correlated. In principle, it is possible to consider a time series as one measurement realization, including temporal correlations in the Bayesian framework. However, this would significantly complicate the framework. An alternative could be to use less samples (e.g. one sample per couple of hours), with the disadvantage of reducing the amount of available measurement data. In the current work, we will simply consider measurements to be uncorrelated in time. This is a conservative approach, since it will lead to an underestimation of the measurement variance, and thus an overestimation of the model error.

\begin{table}[t] \centering 
\caption{Summary of the prior distributions of the empirical model parameters $\boldsymbol{\vartheta}_e$ and statistical parameters $\{\boldsymbol{\vartheta}_t, \boldsymbol{\vartheta}_b \} $ in the Bayesian formulation} \label{tab:priors}
\begin{tabular}{c c c} \hline \hline
$\boldsymbol{\vartheta}_e$ & $\boldsymbol{\vartheta}_t$ & $\boldsymbol{\vartheta}_b$\\ \hline
$k \sim \mathcal{U}(0,1)$ & $\sigma_T \sim \mathcal{U}(0,0.5)$ & $\sigma_{\zeta_i} \sim \mathcal{U}(0,1)$\\
$a \sim \mathcal{U}(0,1)$ & $L_s/D \sim \text{Exp}(0.2)$ & $\delta_{\zeta_i} \sim \mathcal{U}(-1,1)$\\
$b \sim \mathcal{U}(0,1)$ & $L_n = D$ &  \\ \hline \hline
\end{tabular}
\end{table}
 
Computing the joint and marginalized posterior distributions analytically is often not feasible. Therefore a Markov Chain Monte Carlo (MCMC) sampling algorithm is used, as is common practice. In the current work, the posterior probability density functions are approximated with 4 chains of 
1000 samples each obtained via the No U-Turn Sampler algorithm \cite{NUTS} within the Stan \cite{Stan} library in R, i.e. RStan. The burn-in period of each chain is 500 samples. The convergence of the chains and the sampling efficiency in the bulk as well as in the tails of the posterior are validated via the R-hat statistic, the Bulk Effective Sample Size, and the Tail Effective Sample Size respectively.\cite{rhat}

\subsection{Applications to model calibration and comparison}\label{sec:applicationsBI} 

Eventually, once the posterior distributions are known (see above), appropriate parameters must be selected that lead to a well-calibrated version of the wake model, given the data. A common choice is the Maximum A Posteriori (MAP) estimator, as this is the combination of parameters with the highest posterior probability. Other options can be the use of the expected value or the geometric median, but these are not further explored in the current work. Once MAP values $\underline{\boldsymbol{\vartheta}}$ are identified, it is also possible to include an additional bias correction to the model, by simply adding the MAP estimates $\underline{\boldsymbol\mu}_B$ of the model bias to the model predictions. This is however only meaningful, if the proposed Bayesian model is sufficiently adequate (see \S\ref{sec:validation_biframework}).

The Bayesian framework can also be used to compare models proposed in literature by separating the model error distribution and fluctuations in the data. In that case, only the posterior probability density of the statistical parameters is estimated, whereas $\boldsymbol{\vartheta}_e$ is given (e.g. using the MAP estimate $\underline{\boldsymbol{\vartheta}}_e$, or a value obtained from the literature, ...). The likelihood is then given by  
\begin{equation}\label{eq:validation}
\{ \boldsymbol{P}^*_n-\mathcal{M}(\boldsymbol{\vartheta}_e, \boldsymbol{\varphi}^*_n) \} | \boldsymbol{\vartheta}_t, \boldsymbol{\vartheta}_b \sim \mathcal{N}( \boldsymbol\mu_B, \boldsymbol\Sigma_T + \boldsymbol\Sigma_B).  
\end{equation}

\subsection{Validation of the Bayesian framework}\label{sec:validation_biframework}
The Bayesian framework can be validated by computing the distribution of a predicted power measurement $\boldsymbol{\bar{P}}$ for wake model parameters $\boldsymbol{\bar{\varphi}}$  by the Bayesian model $\mathbb{M}$ after conditioning the model parameters $\boldsymbol{\vartheta}$ on the data $\mathcal{D}$. This distribution is also known as the posterior predictive and is given by $\mathbb{P}(\boldsymbol{\bar{P}} |\boldsymbol{\bar{\varphi}}, \mathcal{D}, \mathbb{M})$.\cite{Gelman} If the Bayesian model $\mathbb{M}$ is adequate, the distribution of a predicted power measurement should resemble the distribution of the observed data $\mathcal{D}$. Because of the assumption that the data vectors $\boldsymbol{P}^*_n$ are independent given $\boldsymbol{\vartheta}$, the posterior predictive can be expressed in terms of the likelihood and the joint posterior distribution of $\boldsymbol{\vartheta}$ as
\begin{equation}
\begin{split}
\mathbb{P}(\boldsymbol{\bar{P}} | \boldsymbol{\bar{\varphi}}, \mathcal{D}, \mathbb{M}) &= \int \mathbb{P}(\boldsymbol{\bar{P}}| \boldsymbol{\bar{\varphi}}, \boldsymbol{\vartheta}, \mathbb{M} ) \mathbb{P}(\boldsymbol{\vartheta} |  \mathcal{D}, \mathbb{M})d \boldsymbol{\vartheta}  \\ 
&\approx \frac{1}{M} \sum_{m=1}^{M} \mathbb{P}(\boldsymbol{\bar{P}}|\boldsymbol{\bar{\varphi}}, \boldsymbol{ \vartheta}^{(m)}).
\end{split}
\end{equation}
The integral can be approximated with a sum of $M$ evaluations of the likelihood $\boldsymbol{\bar{P}}|\boldsymbol{\bar{\varphi}}, \boldsymbol{ \vartheta}^{(m)} \sim \mathcal{N}(\mathcal{M}(\boldsymbol{\vartheta}_e^{(m)}, \boldsymbol{\bar{\varphi}}) + \boldsymbol\mu_B, \boldsymbol\Sigma_T(\boldsymbol{\vartheta}_t^{(m)}) + \boldsymbol\Sigma_B(\boldsymbol{\vartheta}_b^{(m)})) $ at each of the $M$ samples of the posterior $\boldsymbol{ \vartheta}^{(m)} \sim  \mathbb{P}(\boldsymbol{ \vartheta} | \mathcal{D})$. Instead of summing all likelihood functions, we generate one sample $\boldsymbol{\bar{P}}^{(m)}$ from the distribution of the likelihood per $\boldsymbol\vartheta^{(m)}$.\cite{PPC_stan} The $M$ samples of $\boldsymbol{\bar{P}}^{(m)}$ then represent the posterior predictive distribution of a measurement $\boldsymbol{\bar{P}}$ for the conditions $\boldsymbol{\bar{\varphi}}$. By following this procedure for every data point $(\boldsymbol{P}^*_n, \boldsymbol{\varphi}^*_n)$, an adequate Bayesian model should regenerate the distribution of the data. Similarly, the distribution of the model error $\mathcal{E}_B$ is found as a posterior predictive distribution based on the posterior distribution of its mean and covariance.

\section{Results and discussion}\label{sec:res&dis}
The joint posterior probability density of all parameters is presented in Section \ref{sec:res_UQ}. Subsequently, the Bayesian model is validated in Section \ref{sec:res_val_BUQ}. Finally,  Section \ref{sec:res_cal} discussed wake model calibration and comparison. In Sections~\ref{sec:res_UQ} and \ref{sec:res_cal} we focus on the $270\pm 2.5^\circ$ wind direction case. A further comparison with the $180\pm 2.5^\circ$ case is included in \ref{sec:res_val_BUQ}.

\begin{figure*}[t]\centering
         \centering
         \includegraphics[width=0.95\textwidth]{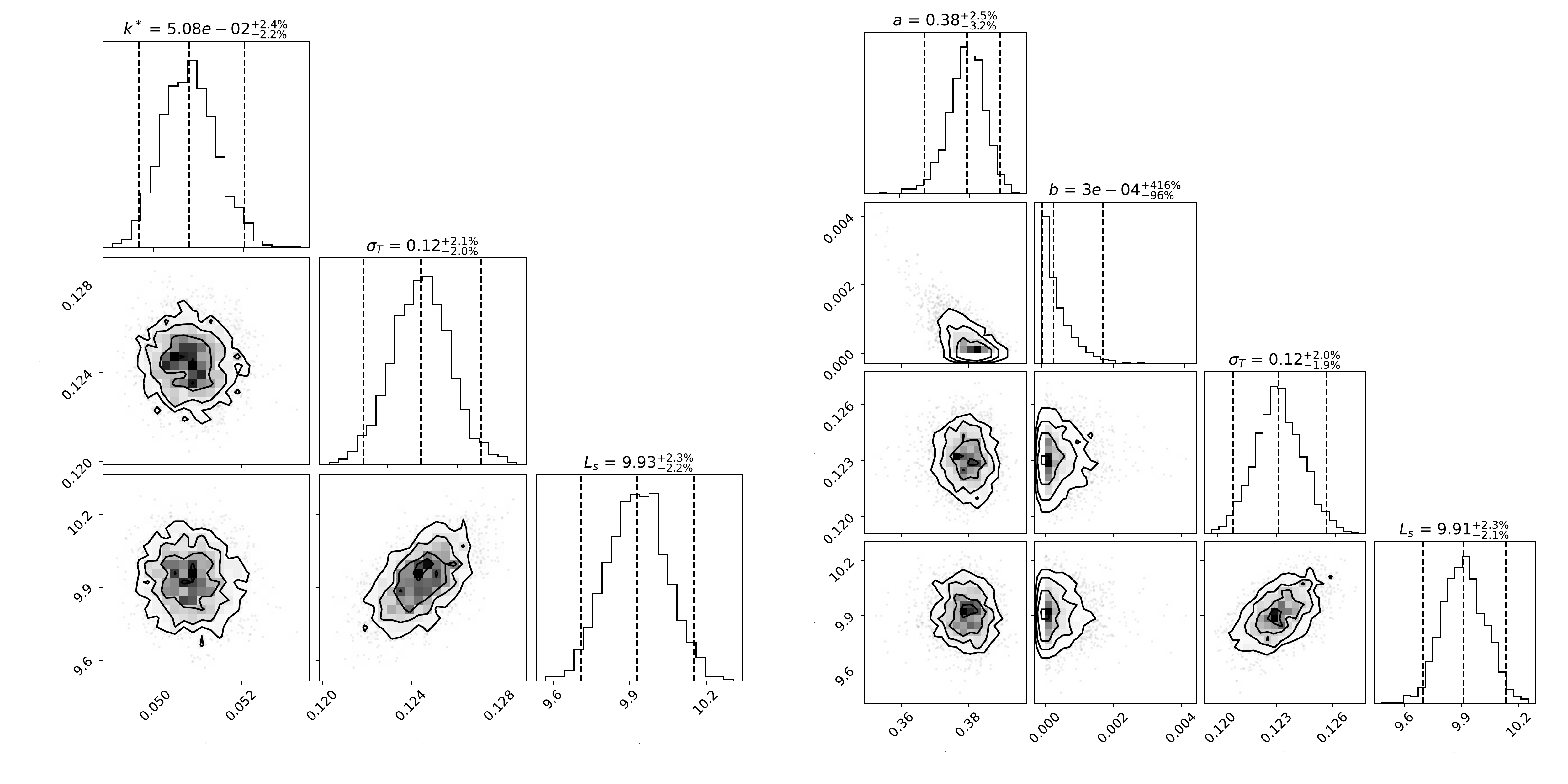} 
     \caption{Joint posterior probability density of the empirical and turbulence-related parameters (270$^\circ$): Constant wake expansion model  (left)  and linear turbulence intensity dependent model (right).\cite{Corner} The dashed vertical lines denote the 0.025, 0.5, and 0.975 quantiles. The titles indicate the values of the medians and the differences with the 0.025 and 0.975 quantiles as percentages of the corresponding median.}\label{fig:posterior}
\end{figure*}

\begin{figure*}[t]
\includegraphics[width=\linewidth]{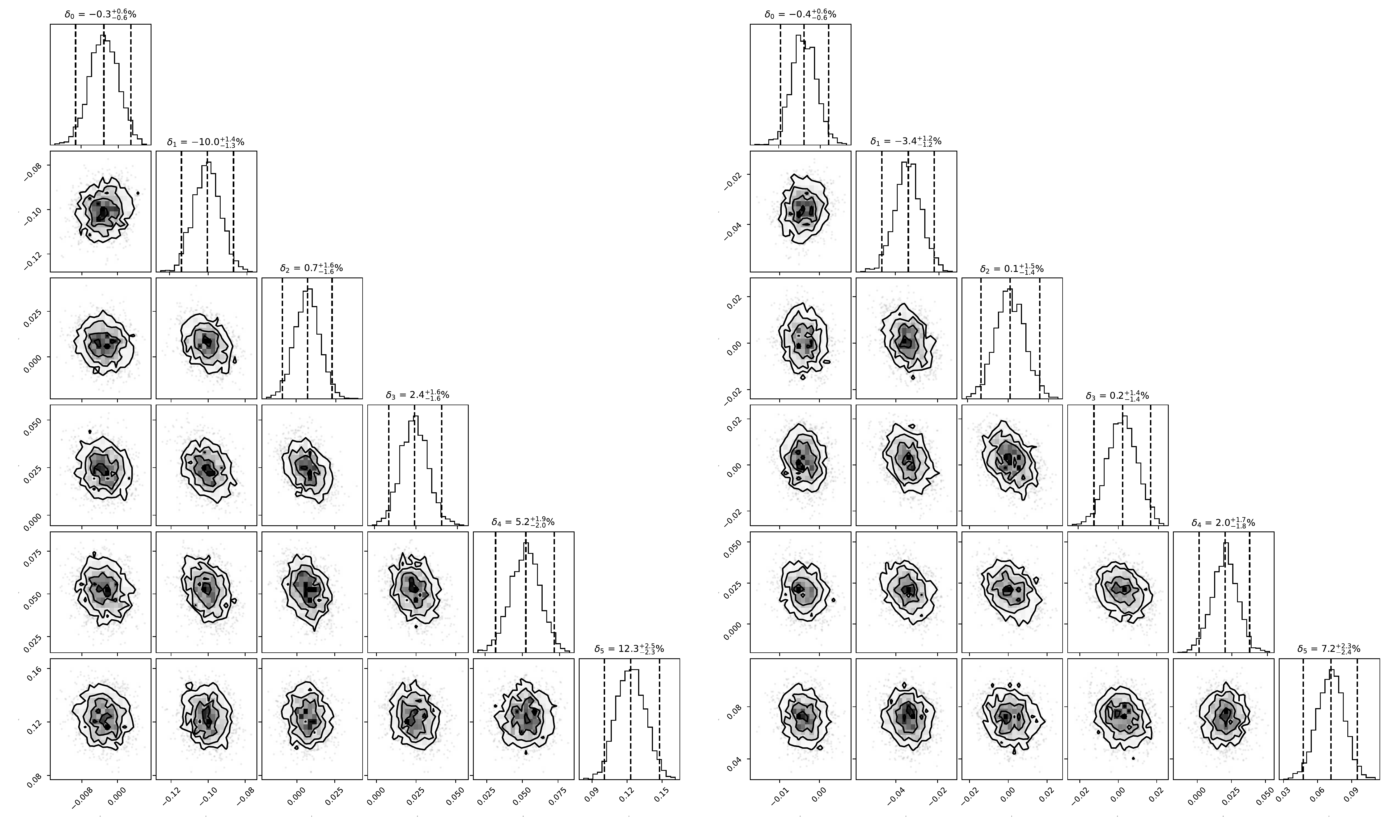} \centering 
\caption{Posterior predictive distribution of the model bias $\boldsymbol{\mu}_B$ (270 $^\circ$): Constant wake expansion rate model (left) and linear turbulence intensity dependent model (right). The bias terms are binned per number of affecting turbines, meaning that $\delta_0$ is the model bias on the upstream turbines \cite{Corner}. The dashed vertical lines denote the 0.025, 0.5, and 0.975 quantiles. The titles indicate the values of the medians and the differences with the 0.025 and 0.975 quantiles, all expressed as percentages.} \label{fig:corner_bias}
\end{figure*}

\subsection{Bayesian uncertainty quantification}\label{sec:res_UQ}
Figure \ref{fig:posterior} depicts the joint posterior distribution of the empirical and turbulence-related parameters for the two wake expansion rate parameterizations, and using data from the $270\pm 2.5^\circ$ wind direction. The medians and 95\% posterior probability intervals of the empirical parameters posteriors agree with literature values. The constant wake expansion rate $k^*$ is identified with high certainty as the distribution has a standard deviation of approximately 2.3\% and is symmetric. On the other hand, $a$ and $b$ have rather skewed distributions with larger relative uncertainty (especially on $b$). This reflects the uncertainty on the values in literature and can be explained by (1) the general fact that more degrees of freedom result in higher uncertainty on the exact combination of values, (2) the dependence of the parameters as they are correlated, and (3) the insignificant contribution of the $b$ parameter to the model as it is two orders of magnitude smaller than $k^*$. The third argument also explains the high relative uncertainty on $b$. Overall, the distributions of the empirical parameters indicate that the framework is able to identify the correct model parameters.

In both models, the turbulence-related parameters have the same posterior distribution. This is exactly what was anticipated as these parameters can only depend on the data. Moreover, they have small standard deviations (less than 3\%) and symmetric distributions. This indicates that the turbulence-related parameters are well-identified and that the proposed model for the turbulent fluctuations is at least adequate. However, the parameters are correlated because of the parameterisation of the covariance matrix $\Sigma_T$. Overall, the turbulent fluctuation model works as anticipated as it is independent of the wake model. In fact, this was made possible by including a model bias form in the Bayesian framework that does vary per model.

Figure \ref{fig:corner_bias} shows the posterior predictive distribution of the model bias $\boldsymbol{\mu}_B$. 
It should be noted that the bias terms do not add up to zero as they have different amounts of associated turbines $\zeta$ (Eq. \ref{eq:constraint_farmpower}). In both models, the bias on upstream turbines $\delta_0$ is approximately zero, which is a consequence of the computation method for the background velocity field. However, the absolute bias is larger for the constant wake expansion model, which is  explained by the ability of the linear model to capture effects of the added turbulence intensity by the first row of turbines which enhances the mixing of downstream turbine wakes. Deeper in the farm, the model bias distribution becomes wider, which may be related to more unmodelled physics becoming important. As anticipated, the bias terms on turbines with different amounts of affecting upstream turbines are independent, indicating that the specified correlation structure is appropriate. Although the epistemic uncertainty in the wake expansion rate parameter $k^*$ was smaller than the epistemic uncertainty on $a$ and $b$, its model bias distribution is slightly wider. Indeed, the linear model captures the variation in turbulence intensity and therefore has smaller standard deviations for each bias term, i.e. less variations due to unmodelled physics.

\subsection{Validation of the Bayesian framework}\label{sec:res_val_BUQ}
\begin{figure*}[t]
\centering
\includegraphics[width=0.9\linewidth]{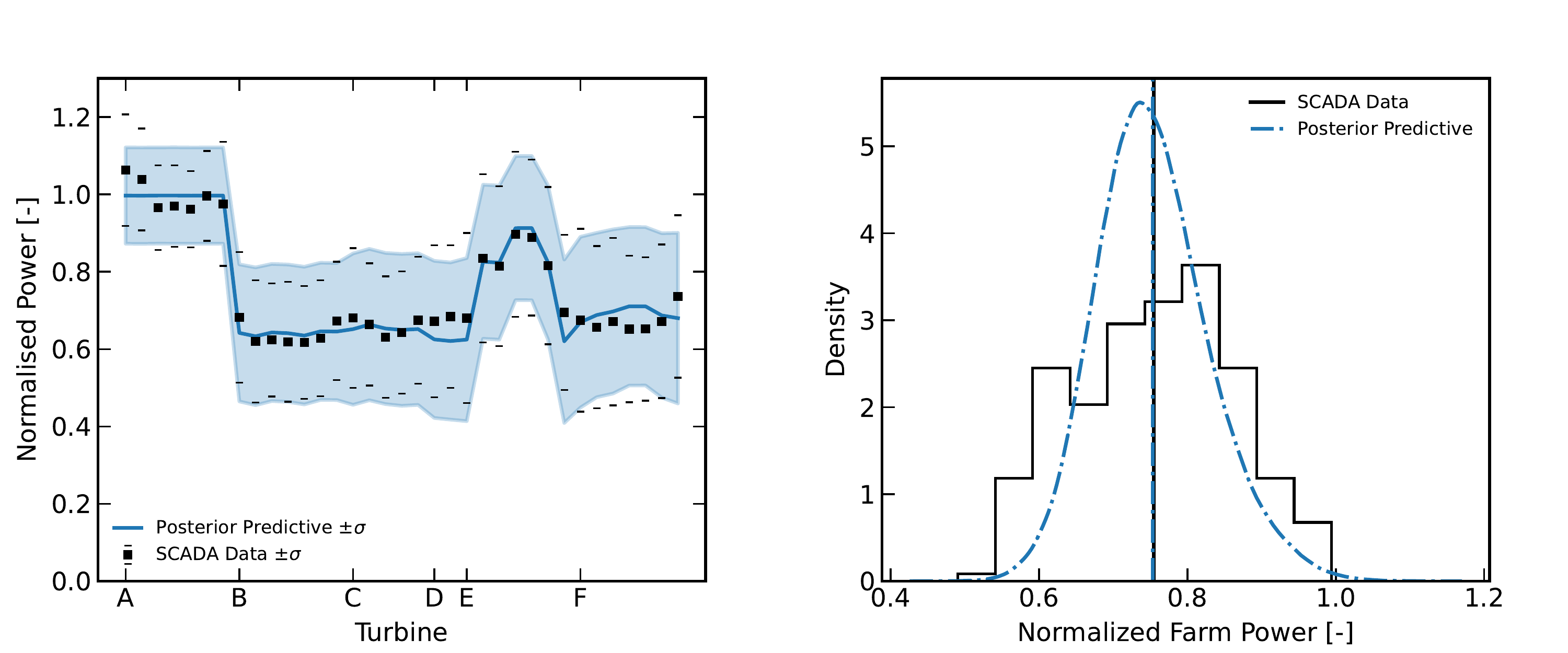} 
\caption{Comparison of the posterior predictive distribution of the power output for the constant wake expansion rate model with the power data, both within one standard deviation (270 $^\circ$): Turbine power (left) and farm power (right). 
} \label{fig:PPC_270}
\end{figure*}

The Bayesian framework is validated in a first step by computing the posterior distribution of the predicted power measurement for the $270\pm 2.5^\circ$ wind direction case, as shown in Figure \ref{fig:PPC_270}. Looking at the mean of the posterior (left panel), we observe that it matches the mean turbine power quite well. Only, in row D and F the posterior mean does not match the mean turbine power. This is explained by the effect of grouping the bias terms per amount of upstream turbines. In fact, the model bias does not only depend on the amount of affecting upstream turbines, but also on the distance between the turbines. For instance, in the current representation, turbines D1 and F5 have the same model bias, but a quite different spacing in their upstream turbines (see Figure~\ref{fig:WMR_layout}). Further looking at the the predicted turbine power variance, we see that it agrees well with the data. Overall, we can conclude that the Bayesian framework quite adequately matches the data distribution's first and second order moments despite the simplified representation of the model errors.

\begin{figure*}[t]
\centering
\includegraphics[width=0.9\linewidth]{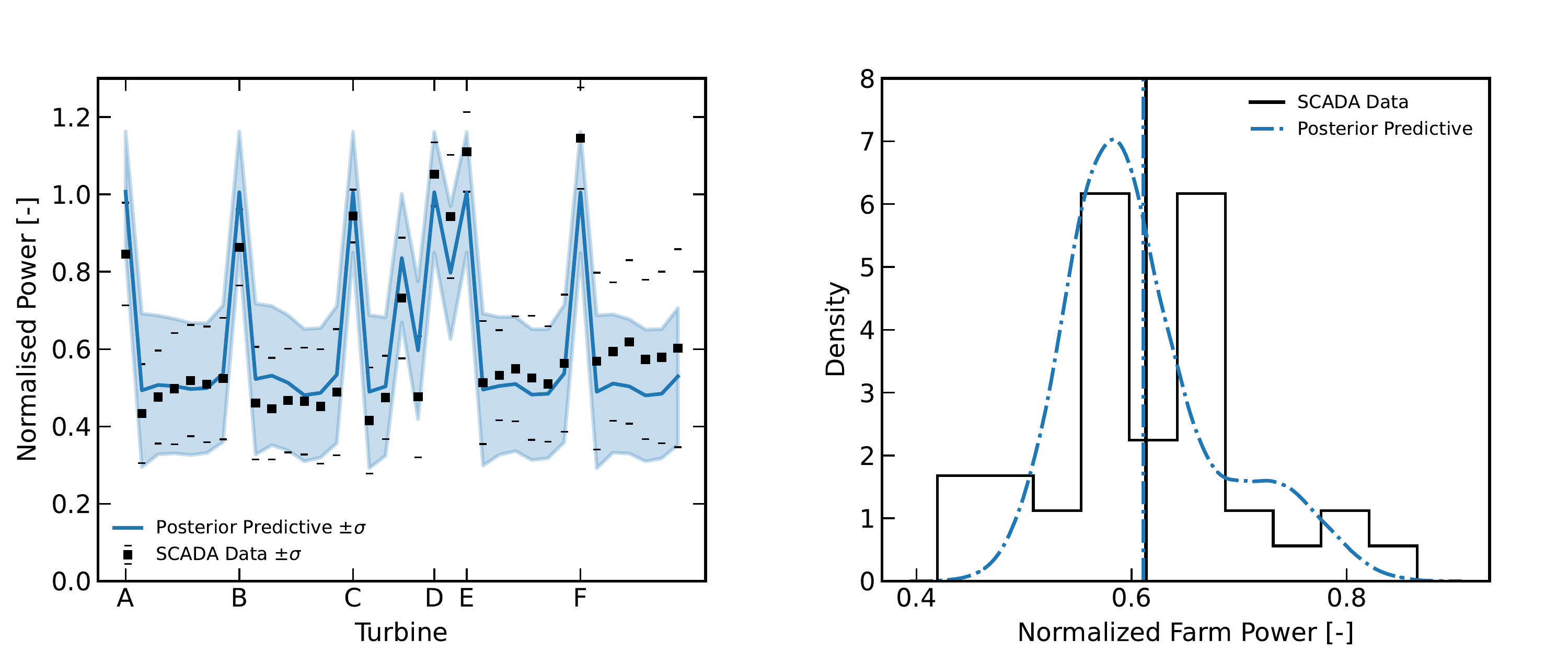} 
\caption{Comparison of the posterior predictive distribution of the power output for the constant wake expansion rate model with the power data, both within one standard deviation (180 $^\circ$): Turbine power (left) and farm power (right). } \label{fig:PPC_180}
\end{figure*}

The posterior predictive distribution of the farm power is given as well in Figure \ref{fig:PPC_270}. The posterior predictive farm power mean matches the mean of the SCADA data well, though they differ slightly by 0.31\%. Indeed, whether the Bayesian modelled mean and mean of the data coincide is not guaranteed by the constraint (Eq.~\ref{eq:constraint_farmpower}), but rather affected by the overall model adequacy. Nonetheless, it should be mentioned that the deviation is small in this case and that it is also partially contributed to uncertainty on the posterior predictive mean due to a limited amount of MCMC samples. 

Further looking at the farm power in Figure \ref{fig:PPC_270} (right), we observe that not all variation in the data is captured, although the comparison between the distributions is quite reasonable. Since the variation in the turbine power does agree with the data (in left panel of Fig.~\ref{fig:PPC_270}), this can be attributed to a possible correlation between the model error of different turbine rows, which is not present in our proposed model error model (cf. Eq~\ref{eq:modelSigma}). This may point to the presence of, e.g., coastal gradients (which are known to exist at Westermost Rough),\cite{Lanzilao} or other unmodelled atmospheric conditions that lead to correlation errors between turbine rows. Clearly, a more general parametrization of the model error that allows to better represent farm-wide correlations, without blowing up the number of parameters in the formulation, is an interesting direction of future research.

In order to verify that the framework works for other wind directions as well, it is applied for a wind direction of $180\pm 2.5^\circ$. MAP values, etc. are quite similar to the $270\pm 2.5^\circ$ case (not further shown), but here we focus again on the evaluation of the adequacy of the Bayesian framework. Figure \ref{fig:PPC_180} shows the posterior predictive distribution for the $180\pm 2.5^\circ$ case. First of all, looking at the left panel, a clear presence of a spanwise gradient in the farm is observed. Looking at the mean power in the data of the first turbines in respectively row A, B, ..., F, (which are the upstream turbines for this wind direction), it is clear that the front of row D, E and F have a significantly higher power then the front of row A--C.  Neither the mean posterior predictive (nor the MAP) follow this trend, since neither the wake model, nor the error model allow for these type of spanwise gradients in the results. Nevertheless, apart from row F, the posterior predictive (both the mean and the variance) corresponds reasonably well with the distribution of the data. Looking at the posterior predictive distribution on the farm level (\ref{fig:PPC_180}, right panel), we observe that it is again reasonably well predicted when compared to the data, with a well predicted mean, and a reasonable prediction of the variance as well.

\subsection{Wake model calibration and model comparison}\label{sec:res_cal}
Given a Bayesian framework that is consistent (cf. discussion above), it is now possible to calibrate the coefficients in the wake model, as well as the model bias. To this end, we use the MAP approach as discussed in \S\ref{sec:applicationsBI}, focusing on the $270\pm 2.5^\circ$ wind direction case. Table \ref{tab:UQ} gives the mean and MAP estimates of the parameter posterior densities. The wake growth parameter distribution agrees with the common value of 0.05 for off-shore wind farms. The coefficients in the linear model also agree with the commonly adopted literature values.

\begin{figure}[t] \centering
\includegraphics[width=0.85\linewidth]{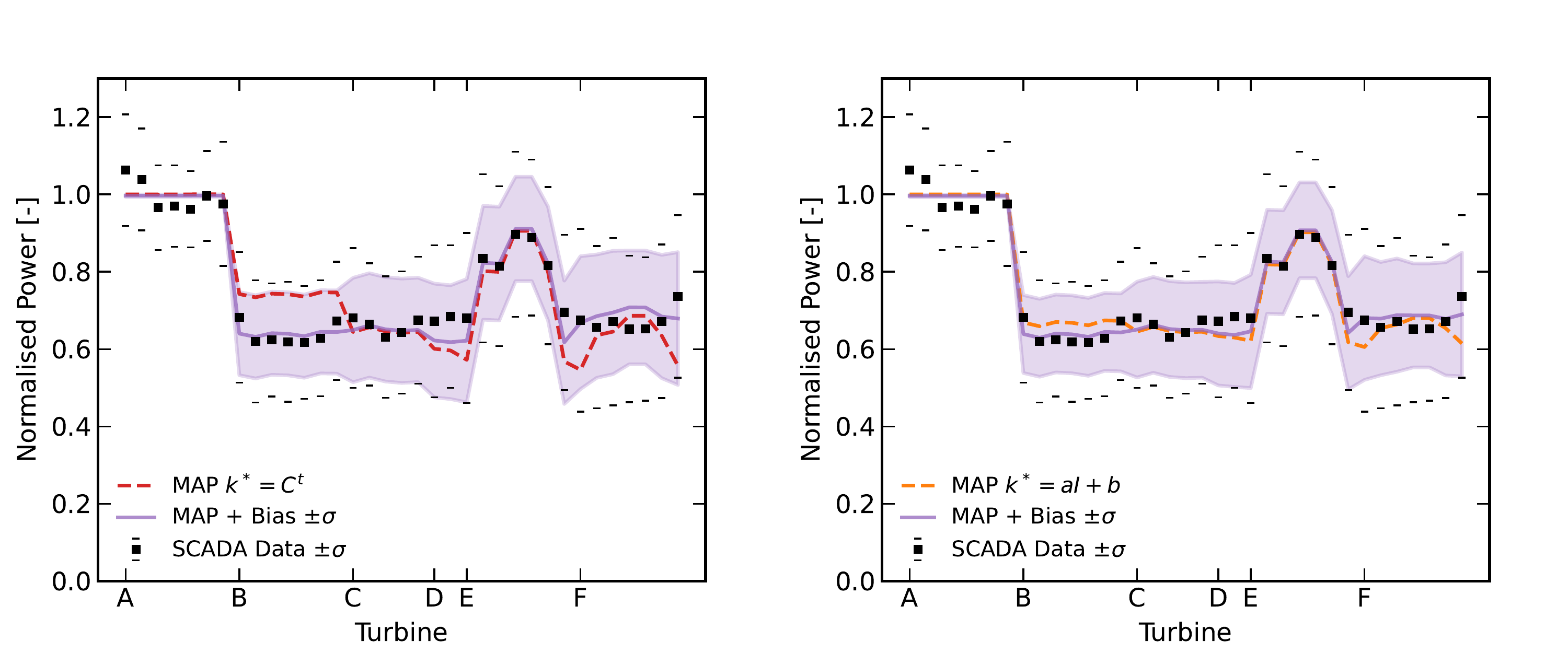} \centering 
\caption{Comparison of the calibrated analytical wind farm power output model with MAP estimates to the power data plus and minus one standard deviation (270$^\circ$). The bias-corrected model (MAP + $\underline{\boldsymbol{\mu}}_B$) plus and minus one standard deviation (from $\underline{\Sigma}_B$) is given as well.} \label{fig:MAP_turblevel}
\end{figure}

\begin{table}[b] \centering 
\caption{Estimated mean and MAP from the posterior distributions of the empirical model parameters for the constant wake expansion rate and the linear turbulence intensity dependent model (270$^\circ$).} \label{tab:UQ}
\begin{tabular}{c c c c} \hline \hline
 & $k^*$ & $a$  & $b$ \\ \hline 
  $\mu$& 0.0508 & 0.3790 & 4.796e-4 \\ 
  MAP  & 0.0508 & 0.3794 & $\approx$ 0 \\ \hline \hline
\end{tabular}
\end{table}

Figure \ref{fig:MAP_turblevel} compares the calibrated models with the data. The figure shows the MAP prediction of both models, but presents as well the bias-corrected MAP prediction and the model-error uncertainty band (see Eq.~ \ref{eq:validation}). Indeed, given a consistent Bayesian framework, we can correct the MAP prediction of the model with the MAP values for the bias terms $\underline{\boldsymbol\mu}_B$. From the figure, it is clear that the calibrated constant wake-expansion-rate model clearly underestimates the wake loss in the second row, but manages to model the wake recovery in row E. The linear model shows less bias as it is able to capture the smaller wake recovery in row A compared to row B. However, once both models are bias-corrected, the differences in the mean prediction become very small. Only in the last row, we find a noticeable difference caused by the interplay of the different MAP estimates of the wake models and the dependence of the model bias on the interturbine spacing.     

Looking at the bias-corrected model-error band in Figure \ref{fig:MAP_turblevel}, it can be seen that it is much smaller than the variance in the data. This simply follows from the fact that the total variance in the data corresponds to $\boldsymbol\Sigma_T + \boldsymbol\Sigma_B$ (see Eq.~ \ref{eq:likelihood}), whereas here only the model error variance is shown. It is one of the strengths of the Bayesian framework to isolate the model error variance in this way. Note that the model error variance in the first turbine row is nearly zero. This is a direct result of the way in which the power is normalized in the model and the data (cf. \S\ref{sec:sources_uncertainty}).

\begin{figure}[t] \centering
    \includegraphics[width=\linewidth]{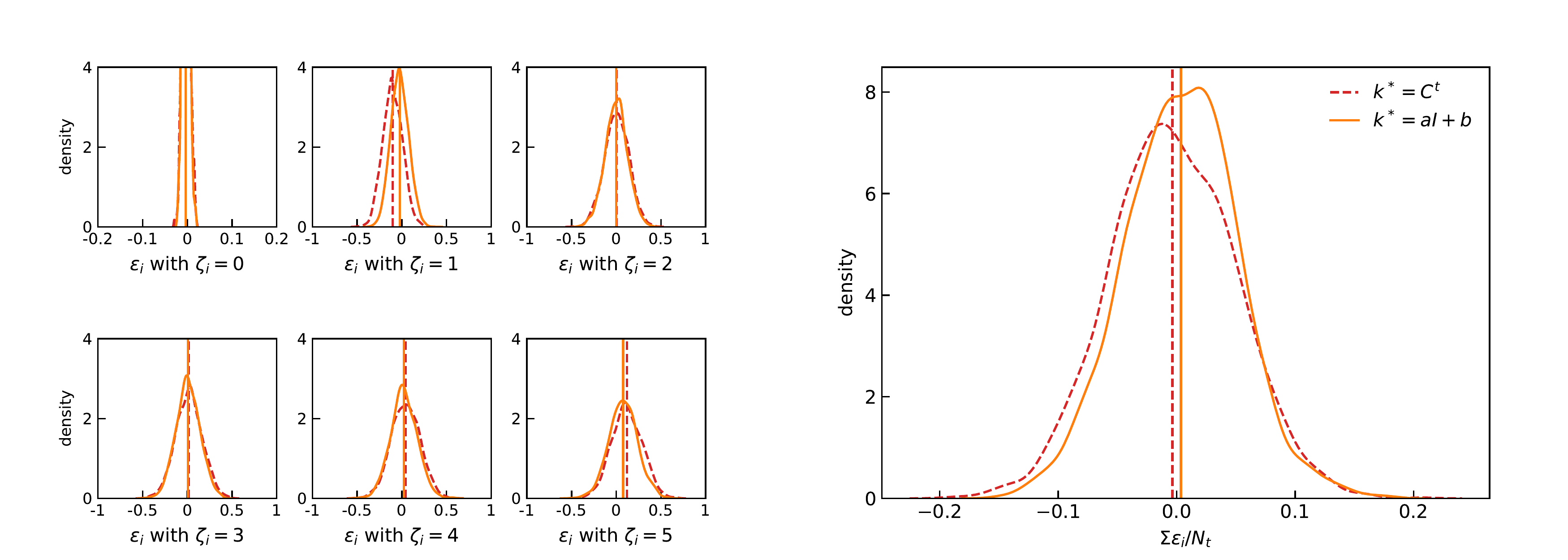}
    \caption{Comparison of the model error distribution (posterior predictive of $\mathcal{E}_B$) for the MAP estimates for a constant wake growth rate and a wake growth rate that is linearly dependent on the turbulence intensity (270$^\circ$). Left: Model error binned per number of affecting turbines $\zeta$. Right: Model error on the farm power.}
    \label{fig:results_validation}
\end{figure}

Figure \ref{fig:results_validation} presents a comparison of both calibrated wake models in terms of their model error distribution, which depends both on the estimated bias $\delta_{\zeta}$ and standard deviation $\sigma_\zeta$. In both models (left panel), the bias is negative for turbines in the second row (i.e. $\zeta_i = 1$ or turbines with one affecting turbine), while it grows for more downstream turbines as mentioned before. The linear model again shows smaller model bias. However, since both models can be corrected with their MAP bias estimates, it may be more important to look at the model variance. We see that the constant wake-expansion-rate model has a slightly higher model error variance than the linear model (as seen in the slightly lower peaks in the distribtions), but this difference is rather small. This suggests that other, unparametrized, physical quantities play a much larger role in the wake model error than the role of turbulence intensity in the wake expansion coefficient. However, further testing over larger data sets and multiple windfarms is needed to corroborate this finding.

Finally, we briefly discuss the possible relation between model error on the farm power (see Figure~\ref{fig:results_validation}) and eventual uncertainty on AEP estimates obtained with a wake model. First of all, before being able to perform full AEP estimates, the model should be trained over multiple wind directions and possibly wind farms. In this case, error distributions such as in Figure~\ref{fig:results_validation} would then provide the expected model error distribution given any selected input condition $\boldsymbol{\varphi}$. To arrive at predictions on the AEP, the distribution $\mathbb{P}(\boldsymbol{\varphi})$ over the farm lifetime should be sampled. However, estimating the model error uncertainty poses additional challenges, since model errors are not guaranteed to be independent between samples. As shown in Eq.~(\ref{eq:modelerrordef}), the model error distribution $\mathbb{P}(\mathcal{E}_B|\boldsymbol{\varphi})$ is in fact a complex projection/transformation of the distribution $\mathbb{P}(\boldsymbol{\psi}')$ of unrecognized physical conditions, which are not guaranteed to be independent between samples (e.g. winter conditions versus summer conditions). In addition, the hidden distribution $\mathbb{P}(\boldsymbol{\psi}')$ in the training data may differ from the actual distribution in the use case. Thus additional research is needed to make a strong link between the model error as observed in  Figure~\ref{fig:results_validation} and uncertainties on AEP estimates. This may include the use of moving block bootstrapping, or hierarchical inference models,\cite{BUQ_hier} and will require as well the use of out-of-sample validation as a means to estimate the potential effects of changes in the hidden distribution $\mathbb{P}(\boldsymbol{\psi}')$.

\section{Summary and outlook}\label{sec:conclusion}
In this article a Bayesian uncertainty quantification framework is proposed for wake model validation and calibration. The framework is applied to two wake models: one that uses a constant wake expansion rate, and a second in which the wake expansion depends linearly on the  turbulence intensity. Historical data of the Westermost Rough farm is used.  The uncertainty on the wake model parameters, the uncertainty in the data due to turbulent fluctuations and the model error distribution are identified by proposing a model with additional parameters for the fluctuations and for the bias. After constructing the complete Bayesian framework with adaptions for calibration, the joint posterior distribution of the empirical wake model parameters, turbulence-related parameters and model bias parameters is approximated.

The posterior distribution of the empirical parameters in the wake model agrees with literature values. The distribution of the constant wake expansion rate is symmetric and has low epistemic uncertainty whereas the parameters in the linear model (the offset especially) show large uncertainty as well as skewed and correlated distributions. 
The joint posterior distribution of the turbulence-related parameters is in both cases well-identified and the same for both models, which must be the case as the fluctuations are independent of the model. On the other hand, the model bias differs. The absolute bias is smaller for the second model as it takes enhanced wake mixing into account via the linear dependency on turbulence intensity. Moreover, the spread of the model bias is larger in the constant wake expansion rate model as it is not able to capture the variations in the data caused by variations in the turbulence intensity. However, this effect is small, which seems to indicate that `unmodelled' physics (such as,  e.g., the coastal gradient that is know to exist at Westermost Rough), introduces large model uncertainty, that dominates subtleties in the wake expansion parameterization.

In summary, it is found that the Bayesian uncertainty quantification framework allows for (1) wake model calibration via maximum a posteriori estimators and inspection of the model adequacy via the posterior distribution, and (2) model validation by separating the model error and measurement error, as well as indicating the effect of unmodelled physics in the distribution of the model bias. For this reason, we believe that this approach can be used for data-driven enhancement of physics-based models.

Further research may apply the framework to other analytical wake models, such as the Ishihara wake model or a wake expansion rate model that depends on more parameters such as the stability class. Another option is to further improve and extend the framework. A future application could include a heterogeneous background flow. However, that will require more data (coming from either measurements or mesoscale models) to distinguish the wind direction, turbulent fluctuations and gradient in the background velocity field. When using data of multiple wind farms or combining LES and historical data, the advantages of hierarchical Bayesian inference can also be exploited.\cite{BUQ_hier}
Further research may also include a measurement error model on the physical conditions such as the yaw offset and the procedure of Gaumond to average the wake model over a bin of wind directions,\cite{Gaumond} or to use all flow directions at once.
Finally, further research into an improved parametrization of the model error, that remains sufficiently low dimensional, and establishing a strong link between model error and AEP uncertainty, is also of interest.

\section*{Acknowledgments}
The authors would like to thank \O{}rsted for access to the SCADA data of the Westermost Rough wind farm, and thank Dr.~Nicolai~G.~Nygaard for useful comments and suggestions on a first draft version of the manuscript. 

\subsection*{Author contributions}
FA, LL, and JM jointly defined the scope of the study. FA developed the Bayesian inference framework, implemented necessary algorithms and performed all simulations. LL provided support for the wake-merging method, and data cleaning. The manuscript was written by FA and JM, and edited by LL.

\subsection*{Conflict of interest}

The authors declare no potential conflict of interests.



\end{document}